\documentstyle[12pt]{article}
\begin{document}

\evensidemargin=-0.5cm
\oddsidemargin=-0.5cm
\setlength{\baselineskip}{6mm}

\begin{flushleft}
{\it Yukawa Institute Kyoto}
\vspace*{-8mm}
\begin{flushright}
YITP-97-34\\
June 1997
\end{flushright}\end{flushleft}
\begin{center}
\vspace{0.6cm}

{\Large\bf Gap Equations from Fermionic Constraints\\
 on the Light-Front}\footnote{To appear in Progress of Theoretical Physics.}\\
\vspace{0.5cm}

{\large\sc Kazunori Itakura}\\

\vspace{0.5cm}

{\sl Yukawa Institute for Theoretical Physics, Kyoto University,
Kyoto 606-01}\\
{\tt itakura@yukawa.kyoto-u.ac.jp}\\

\vspace{0.5cm}

{\bf Abstract}\\

\vspace{0.2cm}
\begin{minipage}{14cm}
\setlength{\baselineskip}{4mm}
{\small We investigate how the gap equations are obtained in the 
light-front formalism within the four-Fermi theories.
Instead of the zero-mode constraint, we find that the ``fermionic constraints" 
on nondynamical  spinor components are responsible for symmetry breaking. 
Careful treatment of the infrared divergence is indispensable for
obtaining the gap equation.}\\
\end{minipage}
\end{center}

\setlength{\baselineskip}{6mm}

\vspace{0.5cm}


\noindent{\bf 1. Introduction.}
Development of our understanding of broken phase physics 
   in the light-front (LF) formalism have been somewhat one-sided 
   with scalar models.
The idea that the spontaneous symmetry breaking in a scalar model 
   is described by solving a constraint on a longitudinal 
   zero mode (ZM), i.e. the ZM constraint \cite{MasYama},  
   has been examined by several authors for
   simple scalar models \cite{Heinzl}-\cite{O(N)}.  
Compared with such extensive works,
   the {\it dynamical} symmetry breaking (DSB) in 
   fermionic systems is not yet sufficiently understood at present.
Especially, the four-Fermi theories which are typical examples
   of DSB should be investigated more minutely before we study 
   the chiral symmetry breaking in QCD.
Absence of bosonic degrees of freedom in the four-Fermi theories
   means that we cannot follow the same way above.
Then, how can we formulate DSB analogously to the idea of ZM
constraints in scalar models?
In this Letter, we consider this problem by viewing how the gap
   equations appear in the four-Fermi theories defined by 
\begin{equation}
{\cal L}=\bar\Psi_a(i \partial\!\!\!/ -m_0 )\Psi_a +
\frac{g^2}{2}\left\{(\bar\Psi_a \Psi_a)^2 + 
                    \lambda(\bar\Psi_a i\gamma_5 \Psi_a)^2\right\},
\ \ \ a=1,\ldots,N.\label{lag}
\end{equation}
Here, we consider an $N$ component fermion $\Psi_a$ and include
   the bare mass.
In the 3+1 dimensional case, the Lagrangian (\ref{lag}) stands for the 
   Nambu-Jona-Lasinio (NJL) model \cite{NJL}
   ($\lambda=1$), whereas in 1+1 dimensions, the chiral Gross-Neveu (GN) model
   ($\lambda=1$) and the non-chiral GN model ($\lambda=0$) \cite{Gross_Neveu}. 
Our standard knowledge in the equal-time (ET) formulation is as follows.
When the bare mass is zero, the (chiral or discrete) symmetries 
   of the systems 
   break down spontaneously due to $\langle \bar\Psi \Psi\rangle \neq 0$.
This is a result of an analysis of the gap equation, which is
   essentially the self-consistency condition on the fermion self energy.
It gives the information of the phase transition as well as the
   nonzero value of $\langle \bar\Psi\Psi \rangle$.
Therefore, it is important to see how the gap equations emerge 
   in the LF formalism.
Despite several works on the LF four-Fermi 
   theories \cite{Heinzl_NJL}-\cite{Maedan2}, however, 
   it is not yet so clear how the gap equations are obtained 
   exactly on the light-front frame without auxiliary fields 
   included.

In the next section, we consider the implication of a nonlinear constraint 
   on the nondynamical component of the spinor 
   (called the ``fermionic constraint") and insist that 
   solving the fermionic constraint is the key to the broken phase description.
In Sec.~3, it is shown that the fermionic constraint 
   in the leading order of $1/N$ expansion indeed becomes the gap equation.
Discussion is given in the final section.\\

\vspace{-0.2cm}

\noindent{\bf 2. Implication of the Fermionic Constraints.} \ 
One of the most outstanding features of the LF four-Fermi theories 
   is the complexity of the fermionic constraints.
In the LF formalism,  half of the  spinor field is 
   a dependent variable and there always exists a constraint on it.
While it is usually solved without any effort, 
   the constraints in the four-Fermi theories are nonlinear
   and thus are difficult to solve.
For example, one of the Euler-Lagrange equations of the GN model 
   is given by\footnote{
Our notation is as follows.  
In 3+1 (1+1) dimensions, the light-front  coordinates  
   are $x^{\pm}=(x^0\pm x^3)/\sqrt{2}$ and $x_\perp^i=x^i$ for $i=1,2$ 
   ( $x^{\pm}=(x^0\pm x^1)/\sqrt{2}$ ) 
  and the derivatives $\partial_{\pm}=\partial/\partial x^{\pm}.$
We use the following two-component representation for the $\gamma$ matrices
$$
\gamma^0=\pmatrix{
0 & {\bf 1}\cr
{\bf 1} & 0\cr
}, \  \ 
\gamma^3=
\pmatrix{
0 & -{\bf 1} \cr
{\bf 1} & 0 \cr
},\ \ 
\gamma^i=
\pmatrix{
-i\sigma^i & 0 \cr
0 & i\sigma^i \cr
}
$$
  ($\gamma_0=\sigma_1,\gamma_1=i\sigma_2$ and 
   $\gamma_5=\gamma^0\gamma^1=\sigma_3$ in 1+1 dimensions). }
\begin{equation}
i\partial_- \chi_a=\left \{ \frac{m_0}{\sqrt{2}} 
- \frac{g^2}{2}(\psi^{\dagger}_b\chi^{}_b+\chi^{\dagger}_b\psi^{}_b)\right \}\psi_a,
\label{constraint}
\end{equation}
where 
$\Psi_a=2^{-1/4}(   \psi_a , \chi_a  )^{T}$.
This includes only 
spatial derivative $\partial_-$ and thus is a constraint.
This is the fermionic constraint.
It is this complexity that makes the analysis of 
   the LF four-Fermi theories very difficult.

Now let us consider the implication of the fermionic constraint.
We should notice that 
   the complexity of the fermionic constraint is not just a difficulty  
   but rather a key to a broken phase physics.
We discuss this through two observations below. 
   
Let us first survey the Yukawa approach for the GN model
   of Ref.~\cite{Maedan2}. 
The GN model can be defined as a limit of the Yukawa-like theory.
At the limit, a scalar field becomes an auxiliary field 
   $\sigma=g\bar\Psi\Psi$ and we recover the GN model. 
In Ref.~\cite{Maedan2}, nondynamical degrees of freedom (the ZM of
   $\sigma$  and the lower spinor $\chi$) were eliminated 
   by solving two coupled equations (the ZM and fermionic constraints) 
   step by step.
Inclusion of the auxiliary field made it easy to solve 
   the fermionic constraint, and essentially the ZM constraint 
   became the gap equation in this approach.
This method leads us to the following conjecture:
   even without auxiliary field included,
   solving only the fermionic constraint
   may correspond to solving the above two constraints.

The second observation is on symmetry. 
Since the lower component of the spinor $\chi$ is a dependent field 
   to be expressed by $\psi$, 
   any transformations should be imposed only on $\psi$.
On the other hand, transformation of $\chi$ is not 
   specified until we solve the fermionic constraint.
So we cannot check whether the models are symmetric or not under 
   the transformations. 
 From this fact, it is reasonable to consider 
   that if we solve the constraint appropriately, 
   we can obtain $\chi$ with some definite transformation property.
Since a $perturbative$ solution turns out to be a symmetric one,
   we must use some nonperturbative method to solve the fermionic constraint.

 From these considerations, we expect that, 
   {\it without auxiliary fields, solving the 
   fermionic constraint nonperturbatively
   must provide us with a description of dynamical symmetry breaking}. 
As is usually done in the ET formalism \cite{Gross_Neveu},   
   the 1/$N$ expansion will be useful for obtaining 
   nonperturbative and nontrivial solutions.

To this end, we set up the problem as follows.
First, we work in infinite space, that is,  do not put the system 
   in a finite box.
This is because the important quantity is not the ZM of $\Psi$ but the ZM
   of the composite operator $\bar \Psi\Psi$. 
Most of it is made of nonzero modes of $\Psi$.
Second, we rewrite the Euler-Lagrange equations entirely with the 
   bifermion operators 
   since the fermion condensate is given as the vacuum expectation value
   of the fermion bilinear.
We introduce two kinds of U($N$) singlet bilocal operators at the equal LF time.
For the GN models, 
\begin{equation}
{\cal M}(x^-,y^-)
=\sum_{a=1}^N \psi^{\dagger}_a(x^-,x^+)\psi_a(y^-,x^+),\  
{\cal C}(x^-,y^-)=\sum_{a=1}^N \psi^{\dagger}_a(x^-,x^+)\chi_a(y^-,x^+),
\end{equation}
and similarly for the NJL model 
  (${\cal M}_{\alpha\beta}$ and ${\cal C}_{\alpha\beta}$ where 
  $\alpha,\beta=1,2$ are the spinor indices).
For example, the fermionic constraint (\ref{constraint}) in the GN model
   is rewritten as
\begin{equation}
i\frac{\partial}{\partial y^-}{\cal T}(x,y)=\frac{m_0}{\sqrt{2}}
\Big({\cal M}(x,y)-{\cal M}(y,x)\Big)-\frac{g^2}{2}
\Big({\cal M}(x,y){\cal T}(y,y)-{\cal T}(y,y){\cal M}(y,x)\Big)
, \label{constraint2}
\end{equation}
where ${\cal T}(x,y)={\cal C}(x,y)+{\cal C}^\dagger(x,y)$ and we
   omitted suffices of $x^-$ and $y^-$. 
We define the theory by these equations 
   with this ordering\footnote{It is almost hopeless to discuss the 
      ordering ambiguity at this stage.
      This is because we cannot define the ordering of $\psi$ and 
      $\chi$ unless we solve the fermionic constraint.} 
   and the usual quantization condition on the dynamical fermions.
This setup is the same as in Ref.~\cite{Thies_Ohta}, 
   which treated the GN models.
However, their description of the broken phase did not resort to the 
   gap equation, which means that it is not clear whether 
   their formulation can predict 
   the existence of the phase transition (though we are always 
   in the broken phase, in fact, in the GN model up to leading order of $1/N$).
Moreover, their method cannot be applied to the NJL model.
Instead, we search for an alternative description 
   which includes the gap equation and the physics of the phase transition.\\

\vspace{-0.2cm}

\noindent{\bf 3. The Gap Equations.}
Let us expand the bilocal operators as
\begin{equation}
{\cal M}(p,q)=
   N\sum_{n=0}^{\infty}\left(\frac{1}{\sqrt{N}}\right)^n\mu^{(n)}(p,q),
\end{equation}
and so on. 
Here ${\cal M}(p,q)$ is the Fourier transformation\footnote{Note that 
   $p,q$ can take negative values because we define 
   the Fourier transformation by 
   $$ {\cal M}(x,y)
   =\int_{-\infty}^\infty \frac{dpdq}{2\pi}e^{-ipx}e^{-iqy}{\cal M}(p,q). $$
 } of ${\cal M}(x,y)$.
Systematic 1/$N$ expansion of ${\cal M}(p,q)$ is given by 
   the boson expansion method 
   which is a familiar technique in the many-body physics \cite{itakura}. 
Among various ways of the boson expansions, 
   the Holstein-Primakoff type is the most useful  
   for large $N$ theories.
For example, the first two orders is given by
   $\mu^{(0)}(p,q)=\theta(p)\delta(p+q)$ and $\mu^{(1)}(p,q)
   =B(q,p)\theta(p)\theta(q)+B^\dagger(-p,-q)\theta(-p)\theta(-q)$,
   where $B(p,q)$ is a bosonic operator satisfying $[B(p_1,p_2), 
   B^\dagger(q_1,q_2)]=\delta(p_1-q_1)\delta(p_2-q_2)$ and so on.
Rewriting the 1/$N$-expanded constraint equation in terms of 
   bosonic operators, we can, in principle, 
   solve the equation order by order and
   find ${\cal C}(p,q)$ expressed by $B$ and $B^\dagger$.

By the way, if the lowest order equation has a nontrivial solution,
   it gives rise to a physical fermion mass 
   $ M\equiv m_0-g^2\langle \bar\Psi\Psi\rangle $.
Using $M$ and $\mu^{(0)}$, 
   the lowest order fermionic constraint is reduced to ($g_0^2=g^2N$)
\begin{equation}
\frac{M-m_0}{M}=\frac{g_0^2}{2\pi}
\int_{0}^{\infty}\frac{dk^+}{k^+},
\label{lowest_constraint_2}
\end{equation}
for the chiral and non-chiral GN models and 
\begin{equation}
\frac{M-m_0}{M}=\frac{2g_0^2}{(2\pi)^3}
\int_{0}^{\infty}\frac{dk^+}{k^+}\int_{-\infty}^{\infty} dk^1dk^2,
\label{lowest_const_NJL}
\end{equation}
for the NJL model.\footnote{Heinzl et al. \cite{Heinzl_NJL}
   obtained the same equation in the mean field approximation.
But conceptually their method is a little different from ours 
   because they substantially obtained it from the self-consistent 
   evaluation of the equation $M= m_0-g^2\langle \bar\Psi\Psi\rangle$ 
   which is the same procedure as in the ET formalism, 
   whereas we found that the fermionic constraint itself became
   eq.~(\ref{lowest_const_NJL}) and $M= m_0-g^2\langle\bar\Psi\Psi\rangle$
   was used just for substitution.
Moreover, they cannot go beyond the mean field approximation, 
   whereas our formulation 
   can solve the fermionic constraint in any higher order. 
}
Here we chose $\langle \bar \Psi i\gamma_5 \Psi \rangle=0$ 
   for the chiral GN model and the NJL model.
If we set $m_0=0$, these equations seem to become ``{\it independent of} $M$''
   and thus these are not the gap equation as they are.
However, this observation is not correct 
   because they are not well-defined until the divergent integral 
   is regularized.
Indeed, by carefully treating the infrared divergences, 
   these equations become the gap equations
   which give nonzero values for $M$ even in the $m_0=0$ case.

Essentially the same equation as eq. (\ref{lowest_constraint_2}) 
   was obtained in Ref.~\cite{Maedan2} 
   as the result of the ZM constraint for the auxiliary field.
There, the gap equation was derived by using the damping factor 
   as an infrared regulator.
Instead of it, let us here introduce some IR cutoff.
As was explicitly pointed out in Ref.~\cite{Maedan2}, we obtain the correct 
   gap equation if we follow the same cutoff schemes as
   those of the standard methods in ET formulation such as the 
   covariant four-momentum cutoff.
Indeed, in Ref.~\cite{Heinzl_NJL}, a noncovariant (rotational
   invariant) three-momentum cutoff was performed to obtain the known result.
But such cutoff seems very artifitial as a LF theory and we here propose 
   another cutoff scheme, {\it the parity invariant cutoff}.
Usually, it is natural and desirable to choose a cutoff so as to preserve 
   the symmetry of the system as much as possible.
For the LF coordinates $x^\pm,\ x^i_\perp$, it would be 
   natural to consider the parity transformation 
   ($x^+\leftrightarrow x^-,\ x^i_\perp\rightarrow -x^i_\perp$) 
   and the two-dimensional rotation on the transverse plane.
In fact, the parity invariance is not manifest
   because the LF time $x^+$ is treated distinctively 
   in the usual canonical fomulation.
However, we find it useful for obtaining the gap equation.
First consider the 1+1 dimensional case.
In momentum space, the parity transformation
   is just the exchange of $k^+$ and $k^-$ and therefore
   the parity invariant cutoff is given by $k^\pm<\Lambda$.
Using the dispersion relation\footnote{It should be noted that 
   the use of the constituent mass $M$ in the dispersion relations 
   corresponds to imposing {\it self-consistency conditions}.
} $2k^+k^-=M^2$,
   we find that the parity invariant regularization 
   inevitably relates the UV and IR cutoffs: 
\begin{equation}
\frac{M^2}{2\Lambda}<k^+<\Lambda.
\end{equation}
In 3+1 dimensions, the condition $k^\pm<\Lambda$ gives ($2k^+k^--k_\perp^2=M^2$)
\begin{equation}
   \frac{k_\perp^2+M^2}{2\Lambda}<k^+<\Lambda.\label{PIC}
\end{equation}
Note that this also  means the planar rotational invariance
    $k_\perp^2<\Lambda^{'2}=2\Lambda^2-M^2$.
Since both of the IR cutoffs include $M$,  
   right-hand-sides of eqs. (\ref{lowest_constraint_2}) and 
   (\ref{lowest_const_NJL}) become {\it dependent of $M$}. 
Indeed they are estimated as
\begin{eqnarray}
&&\frac{M-m_0}{M}=\frac{g_0^2}{2\pi}\ {\rm ln} \frac{2\Lambda^2}{M^2},
\label{gap_GN}\\
&&\frac{M-m_0}{M}=\frac{g_0^2\Lambda^2}{4\pi^2}\left\{
2-\frac{M^2}{\Lambda^2}\left( 1 + {\rm ln}\frac{2\Lambda^2}{M^2}  \right)
\right\}.\label{gap_NJL}
\end{eqnarray}
These are the gap equations.
Both equations have nontrivial solutions $M\neq 0$ even in
   the $m_0\rightarrow 0$ limit.
The somewhat unfamiliar equation (\ref{gap_NJL}) of the NJL model 
   shows the same property 
   as the standard gap equations of the ET quantization.
For example, there is a critical coupling $g_{\rm
cr}^2=2\pi^2/\Lambda^2$, above which $M\neq 0$.
In the GN models, we must renormalize the divergence.
The same renormalized gap equation and thus the same
   dynamical fermion mass as the ET result are obtained if we
   renormalize the UV and IR divergences together.
Now it is almost evident that the essential physical role of 
   eq. (\ref{lowest_constraint_2}) is not a `` renormalization 
   condition'' as was discussed in Ref.~\cite{Thies_Ohta},  
   but ``the gap equation''.

Though it was already pointed out in Ref.~\cite{Maedan2}, 
   the following is very significant and worth while to be mentioned again.
The essential 
   step to obtain the gap equation is the inclusion of mass information
   as the regularization rather than the fact that the UV and IR cutoffs 
   are related to each other.
If we regulate the divergent integral without mass information 
   ({\it e.g.} introducing the UV and IR cutoffs independently 
   as in Ref.~\cite{Thies_Ohta}),
   we cannot reproduce the gap equation.
The loss of mass information is closely related to 
   the fundamental problem 
   of the LF formalism \cite{Nakanishi_Yamawaki} and the parity invariant 
   regularization can be considered as one of the prescription for it.
The problem is that the light-front quantization gives a 
   mass-independent two-point function 
   (at equal LF time), which contradicts the result from general
   arguments of the spectral representation.
A symptom of this problem can be easily discovered as absence of mass
   dependence in the usual mode expansion on the LF.
For 1+1 dimensional scalar and fermion fields, 
\begin{equation}
\phi(x)=\int_0^\infty
\frac{dk^+}{\sqrt{4\pi} k^+}[a_{k^+}e^{-ik^+x^-}+a_{k^+}^\dagger
e^{ik^+x^-}],
\end{equation}
\begin{equation}
\psi(x)=\int_0^\infty \frac{dk^+}{\sqrt{2\pi}}[b_{k^+}e^{-ik^+x^-}+d_{k^+}^\dagger
e^{ik^+x^-}],
\end{equation}
with $[a_{k^+}^{}, a_{p^+}^\dagger]=k^+\delta(k^+-p^+), 
\{b_{k^+}^{}, b_{p^+}^\dagger\}=\delta(k^+-p^+)$ and so on.
It is evident that these have no mass dependence and thus need some 
   modification.
In a scalar field, a parity invariant IR cutoff is sometimes 
   introduced \cite{parity_scalar} 
   in order to restore the necessary mass dependence.\\

\vspace{-0.2cm}

\noindent{\bf 4. Discussion.}
We found that the gap equation was derived from the fermionic 
   constraints in the LF four-Fermi theories.
To obtain the gap equation, 
   it was imperatively necessary to include the mass parameters and use 
   the constituent masses  as the IR cutoffs.
This requirement can be understood as a prescription for 
   the common pathology in the LF field theories.
It seems suggestive that we are naturally guided by  
   symmetry such as parity to remedy this problem. 

If we solve the fermionic constraints up to the next leading order and 
   insert the solution into the equations of motion,
   we can derive the fermion-antifermion bound-state
   equations \cite{preparation,itakura_sum}.  
For the GN models,
   they are essentially the same as the results of Ref.~\cite{Thies_Ohta}.
There are solutions for the NG bosons in the chiral limit.
Especially in the chiral GN model, we can obtain an analytic
   solution of the bound-state equation even in the massive case 
   and its mass spectrum near the chiral limit \cite{itakura_sum}.

The way of describing the DSB on LF in our models may be 
   summarized as follows.
By solving the fermionic constraints, which are characteristic of the 
   LF fermionic theories, we have in general two kinds of solution;
   the symmetric and broken solutions.
Substituting these into the Hamiltonians or the eqs. of motion, 
   we reach symmetric or broken theories with trivial vacuum.
Instead of finding new vacuum, the effect of broken phase on the LF
   can be seen as modification of the Hamiltonians.
However, these arguments do not predict which theory should 
   be realized in reality.
This problem will be discussed in more detail elsewhere \cite{preparation}.\\

\vspace{-0.3cm}

The author would like to thank K. Harada, S. Maedan, 
M. Taniguchi, and S. Tsujimaru for valuable comments and suggestions.
Most of this work has been done while the author was staying
University of Tokyo, Komaba.

\end{document}